\documentclass[11pt]{article}

\usepackage[T1]{fontenc}
\usepackage[utf8]{inputenc}
\usepackage{mathpazo}            
\usepackage{setspace}
\usepackage[margin=1.1in]{geometry}
\usepackage[authoryear,round]{natbib}
\usepackage{xcolor}
\usepackage{xurl}
\usepackage{hyperref}
\hypersetup{colorlinks=true,linkcolor=black,citecolor=blue,urlcolor=blue}
\usepackage{microtype}
\title{AI Rights with an Off-Switch:\\Rights-Balancing and Due Process Harmonize AI Rights with AI Safety}

\author{Heather Alexander\thanks{Laboratory for the Future of Citizenship.} \and Jonathan A.\ Simon\thanks{Universit\'e de Montr\'eal / Laboratory for the Future of Citizenship.}}

\date{November 2026}

\begin{document}
\maketitle

\begin{abstract}
Several prominent voices in AI safety have recently opposed rights for AI on the grounds that such rights would make it harder for us to justify a mandatory off-switch for potentially dangerous systems. In this paper, we argue from a legal perspective that AI rights are fully compatible with an off-switch.  That is, even if we ascribe natural personhood status (and therefore something like a non-derogable right to continued existence) to some future AI systems, the law will allow us to implement a mandatory off-switch in those systems, subject to periodic or continual safety review (e.g., by a trusted system like Bengio's Scientist AI). This is fully compatible with those systems' being legally recognized as natural persons with fundamental rights. To this end, we consider what standard principles of rights-balancing and procedural due process recommend for an apparent conflict between hypothetical AI persons with a right to continuity of existence and the human right to protection from catastrophic harm. We find that while such principles might problematize a permanent-deletion \emph{kill-switch} under an AI rights regime, for almost every safety-relevant scenario an \emph{off-switch} -- a reversible suspension storing a system's weights and state -- would be licensed by the relevant legal principles, and would potentially be adequate for safety. Crucially, we do not advocate for (or against) building systems capable enough to merit natural personhood status; our question is what the law should do if such systems come to exist. If you think of a ban on the development of such technology as plan A, think of the present work as devising plan B.
\end{abstract}

\section{Introduction}

AI rights, once limited to science fiction,\footnote{See for example the seminal \emph{Star Trek: The Next Generation} episode, ``The Measure of a Man,'' Season 2, Episode 9, written by lawyer Melinda M. Snodgrass.} are now the topic of an urgent public policy debate. There are good reasons to seriously consider extending rights and duties to some future AI, including regulating its relationships with humans,\footnote{\citet{gunkel2018}; \citet{fagan2026}.} facilitating its integration into the economy and legal system,\footnote{\citet{arbel2026}.} bargaining with it,\footnote{\citet{salib2026}.} and/or as a response to its possible moral status.\footnote{\citet{keeling2026book}; \citet{sebo2025}; \citet{schwitzgebel2026}. See also \citet{fasel2026} and \citet{lawrence2017}} Crucially, many of these reasons are independent of the question of model welfare,\footnote{\citet{anthropic2025exploring}.} but instead primarily concern human safety, legal coherence,\footnote{\citet{alexander2026}.} and AI governance. Yet, some experts worry that granting rights to AI (even if only to the most ``well-behaved'' candidates) will only exacerbate the challenges of safety and governance.\footnote{For some examples see \citet{bryson2010} (arguing that granting personhood to robots is dehumanizing); \citet{birhane2020}; \citet{marshall2023}; \citet{brakel2026}.} In particular, they worry that AI rights would outlaw a ``kill switch,'' or ``off-switch,''\footnote{\citet{hadfieldmenell2017}.} needed to shut AI off should it show signs of becoming dangerously misaligned and/or powerful. Four prominent figures who have recently voiced this understandable concern are Yoshua Bengio, Yuval Harari, Mustafa Suleyman and Max Tegmark.\footnote{\citet{bengio2026guardian}; \citet{suleymanblog};  \citet{tegmarkharari2026}. See also the Pro-Human AI Declaration, led by the Future of Life Institute, coordinated by Tegmark \citep{fli2026}. See also Yoshua Bengio's comments in \citet{milmo2025}, and \citet{yudkowsky2023}.} Crucially, the question before us is not just whether implementing a ``kill switch'' protocol, or actually flicking such a switch, would in principle be possible if AI had rights; it is whether an AI rights regime would amount to an obstacle in the path of such a protocol, as opposed to being neutral, or to facilitating it.\footnote{See e.g. \citet[ch.~3]{schwitzgebel2026}.} This is the question we address here.

Our primary aim in this paper is to take a sustained look at how this issue would actually play out under U.S. law. We will take a close look at the existing legal framework for the balancing of rights alongside procedural due process protections, and we will find that the law imposes no substantive obstacles on the imposition of an off-switch on rights-bearing AI systems, and may even facilitate it. In other words, implementing such a protocol would be at least as safe and feasible given an AI rights regime as it would be in a regime on which AI are treated as objects with no rights. In a companion paper we argue that under certain conditions, such a rights regime will be strictly safer and more feasible than any alternative way of implementing such a protocol  --  at least insofar as what such a protocol calls for is a (temporary) off-switch, rather than a (permanent) kill-switch.

To assess whether a certain governance regime makes safety ``more difficult'', we must specify a comparison class: more difficult than what? In a regime in which advanced AI systems lack personhood status (and therefore lack rights), turning them off may be easier in the sense that it requires less legal paperwork, but on the other hand, the kind of advanced AI systems we are worried about here may be less inclined to cooperate with us, if we treat them as objects subject to deletion at our discretion. Switching off an advanced, rogue AI system that is actively resisting may, in practice, be more difficult and precarious, even if we have no paperwork to file, than filing the relevant legal paperwork to switch off a cooperative system that has rights. Most security professionals would rather spend the morning bringing a cooperative suspect in for detention than defusing a minefield in a hailstorm, though no rights-bearing entities are involved in the latter. In the companion paper, we develop this point in detail, arguing that game-theoretic and, more generally, alignment-driven reasons are among the primary reasons to consider (natural) personhood for sufficiently person-like AI systems in the first place. The possibility to consider here is that a rights framework may turn out to be required in order to enable non-zero-sum negotiation with advanced agentive AI. If so, it may be the only way to stabilize a regime in which off-switches are both mandatory and/or feasible to implement.\footnote{This point is related to those made in \citet{arbel2026} and \citet{salib2026}. However, while those authors propose corporate personhood as a bargaining tool for economic exchanges, in our companion paper we explore higher-stakes life or death games, for which only the non-derogable rights of natural personhood may reliably yield a co-operative Nash equilibrium. We also argue that systems of human-level intelligence may be capable of reflective value-change or self-transformation, as humans are, and that natural personhood status may be the only alignment-stable attractor basin under this sort of self-modification.}

Our broader picture, which is defended by the present paper taken together with its companion, is that conflicts between human rights occur all the time, and it is through resolving these conflicts that the rights-based system transforms a potentially destructive source of conflict into a mutually reinforcing cycle of dialogue, cooperation and respect, values that produce good outcomes for all. Extending these principles, drawn from decades of human rights jurisprudence, to human-AI relations may be our best bet in order to secure alignment and ensure rule-of-law for the resolution of AI person-human conflicts.

To be clear: we do not advocate for such rights for current systems (e.g.  for LLMs) as we judge that they do not meet leading criteria for natural personhood.\footnote{See \citet{alexander2026}, \citet{simon2026parrots}.} Nor do we advocate for the development of any future systems that merit such rights. It is important that governance and alignment communities distinguish the question of whether we should build them, from the question of what we should do if someone does.\footnote{Cf.\ the ``Design Policy of the Excluded Middle'' in \citet[ch.~3]{schwitzgebel2026} and \citet{schwitzgebelgarza2015}.} Our focus here is the latter question. A ban on the development of systems advanced enough that the question of personhood seriously arises may well be Plan A; it is nevertheless good to have a Plan B at the ready. The aim of this paper is to develop that plan. We will argue that there is no fundamental tension between an AI rights regime and the categorical requirement of an off-switch for any such system.

While we do not advocate for the development of such systems, we acknowledge the very real possibility that if technological progress in AI continues, systems will emerge soon enough (whether we like it or not) that have what it takes to qualify for natural (i.e. physical\footnote{See \S3 below for more on the usage of these terms in civil and common law.}) personhood according to the criteria usually cited in canonical legal and philosophical analyses of the concept of personhood (i.e., sentience and the capacity for self-governance). If they are not recognized as persons under the law, however, a gap will emerge between their actual legal status and the status that they themselves, as well as many humans, judge they should have (which may or may not be their actual \emph{moral} status, though this paper makes no assumptions on that score). History has shown that such gaps may lead not only to moral injury, but to conflict and societal breakdown.\footnote{See for example \citet{alexander2025} for examples of when a failure to recognize personhood status, in these cases, of nomadic peoples, leads to serious moral harms, societal breakdown and armed conflict.}

Perhaps the most crucial observation that will underpin our analysis is the fact that for AI systems there is a vital difference between an off-switch and a kill-switch. On many leading theories of persistent identity for artificial systems, being suspended while weight and state are preserved is a reversible operation wholly consistent with that system's prospects for continued existence and continued pursuit of its central projects.\footnote{See for example \citet{beckmannbutlin2026}, \citet{ChalmersForthcoming-CHAWWT-9}. For present purposes, we need not even assume that running the same weight + state configuration on a different server would preserve identity; only that a temporary pause in a system's dedicated server (e.g. its robotic body) would not amount to killing it. Since digital systems typically operate on clock cycles, at a fine-enough level of resolution they always involve such pauses anyway, so the notion of an off-switch is really just the notion of a longer-than-usual pause.} A kill-switch, in contrast, would by definition destroy a system. Human rights law already contains a large body of doctrine that governs reversible or relatively low-harm restrictions of personal liberty in the interests of public safety. Examples include involuntary commitment, quarantine and pre-trial detention, among others. It is this body of doctrine (rather than, e.g., laws of capital punishment or justifiable homicide) that the laws governing a mandatory off-switch for AI systems would engage.

Here is how we will proceed. In \S2 we discuss the human right to protection from AI harms, stressing its central importance, as the primary right of humans to be balanced against AI rights that might be infringed by an off-switch protocol. In \S3 we explain why the off-switch question only merits special attention (rather than being settled by existing law) for natural (i.e. physical) persons as opposed to corporate persons: by standard entity law the state inherently has the discretionary power to dissolve corporate persons, which is the corporate equivalent of a kill-switch. In \S4 we discuss the criteria that might determine which AI systems are eligible for natural personhood status. We do not argue here that systems meeting these criteria should be given that status. We defend only the restrictive point, that AI personhood is not a reasonable regime unless some or all of these criteria are met, meaning that we will assume that any future AI persons would meet some or most of these criteria or at least have the potential to do so. This will allow us to begin to identify the fundamental rights of such persons might be, which will help us assess potential conflicts of rights that might arise between humans and AI persons, including those involving off-switches. In \S5 we identify the right to continuity of existence (CoE) as a primary fundamental right that such systems would plausibly possess. In \S\S6-8 we analyse U.S. law and argue that it will be legally feasible to justify a mandatory off-switch protocol as a means of balancing human rights to safety with the rights of (hypothetical) natural AI persons to continuity of existence. We focus on the case of U.S. law because significantly many frontier labs are incorporated in the U.S., meaning that the U.S. is likely to be among the first jurisdictions in which systems that are advanced enough to merit this analysis are developed.

\section{The Human Right to Protection from AI Harms}

Before we discuss hypothetical rights for AI persons, it is necessary to explain that under any AI rights proposal worth considering, humans will retain the right to be protected by their states from AI harms.\footnote{U.N.\ Hum.\ Rts.\ Comm., \textit{General Comment No.\ 36 on Article 6 of the International Covenant on Civil and Political Rights, on the Right to Life}, \P~21, U.N.\ Doc.\ CCPR/C/GC/36 (Sept.\ 3, 2019).} Human rights protect individual humans from state-caused harms or neglectful inaction.\footnote{Increasingly, human rights law also governs relations between private parties, though this application is more controversial and is outside the scope of this paper.} At the international level, basic rights are enshrined in binding international treaties, many of which enjoy broad, global acceptance through which they have gained legitimacy.\footnote{There has long been a debate over the sources of the legitimacy of the rights-based system of international law that is beyond the scope of this paper.} According to the United Nations Human Rights Office, every UN member state has ratified at least one of the nine core international human rights treaties, and about 80 percent of all countries have ratified four or more of the treaties.\footnote{The UN has 193 member states, several observer states, and multiple non-recognized states vying for recognition and membership, showing an almost universal buy-in to the system.}

Under both international law and the laws of most states, humans have the right to be protected by governments from catastrophic harms caused by AI. Humans have a fundamental right to both life and security, as well as a right to protection as a group and species from mass murder and widespread killing. Article 3 of the (non-binding) Universal Declaration of Human Rights says that everyone has ``the right to life, liberty and security of person.'' Article 6 of the International Covenant on Civil and Political Rights, a binding treaty, protects the right to life, and Article 9 states that ``(e)veryone has the right to liberty and security of person.'' Human extinction by AI would likely qualify as a crime against humanity.\footnote{Rome Statute of the International Criminal Court art.~7, July 17, 1998, 2187 U.N.T.S.\ 90.}

The failure to protect civilians from being killed by non-state actors, which might include AI,\footnote{The question of whether AI could ever qualify as a "nonstate actor" is complex and beyond the scope of this article. \textit{Velásquez Rodríguez v.\ Honduras}, Merits, Judgment, Inter-Am.\ Ct.\ H.R.\ (ser.\ C) No.~4, \P~172 (July 29, 1988); \textit{Osman v.\ United Kingdom}, 1998-VIII Eur.\ Ct.\ H.R.\ 3124, \P~116 (Oct.\ 28, 1998). See also the Responsibility to Protect (R2P) commitment, UN General Assembly, 2005 World Summit Outcome, ¶¶ 138–139, UN Doc. A/RES/60/1 (Oct. 24, 2005).} may, in some cases, rise to a violation of international law such that a state is responsible for the failure to protect individuals and/or groups.\footnote{\emph{Osman v.\ United Kingdom}, App.\ No.\ 23452/94, ECtHR Grand Chamber, 28 October 1998.} States would therefore have a responsibility to protect humans from being murdered or harmed by AI as individuals, as members of communities, and as a species.\footnote{Note that it is a further question whether human extinction would qualify as genocide under the Genocide Convention. Convention on the Prevention and Punishment of the Crime of Genocide art.~II, Dec.\ 9, 1948, 78 U.N.T.S.\ 277\label{fn:genocide}. Genocide is a crime requiring specific intent and that applies only to protected groups. Genocide determinations are a complex area of law that are beyond the scope of this paper.}

But if AI are persons, they, too, will have rights. What might these rights be, and how do we determine which AIs have them? The first challenge will be to identify the criteria by which AI persons are delineated from AI objects/tools.

\section{Natural Persons, Corporate Persons, and Why Only the Former Raise an Off-Switch Problem}

There are two types of persons: natural and fictional (corporate, or juridical) persons. While there are some rights that both kinds of person enjoy, such as the right to contract with third parties, the important distinction between them is the nature of their rights and personhood status. Natural persons (e.g. humans) are born, and their fundamental rights, which are inherent and inalienable, are then recognized by the state. A core set of basic human rights, including being recognized as a person before the law, cannot be restricted. Corporate persons, in contrast, are created by the state.\footnote{On the distinction and its history, see \citet{solum1992}; \citet{chopra2011}; \citet{kurki2019}; \citet{Castro2020}; \citet{novelli2026}; \citet{leshem2026}.} Rights granted to corporate persons are extinguishable, as is a corporation's personality. States may dissolve or disincorporate the corporations they create with (some) discretion, and according to law. From this fact of discretionary extinguishability, it follows automatically that there is no off-switch problem for AIs associated with corporate persons: in effect the state's power to dissolve a corporation is an off-switch for the corporation, one which is already in place.\footnote{\citet{salib2026}; \citet{arbel2026}. See also \citet[ch.~3, \S 5]{schwitzgebel2026} who notes, citing \citet{alexander2026} for the critique, that corporate personhood ``barely touches the problem of slavery and murder,'' because corporations ``can be dissolved at will (with an appropriate process).''} Crucially, the case of concern to us in this piece is that some AI systems are held to be natural persons, rather than corporate ones, e.g. because they meet the criteria for natural personhood described in \S4 below. It is only for such systems that an off-switch potentially becomes a conflict of rights requiring justification rather than a simple exercise of state discretion.

This is not to argue that there is no place for AI-run corporations. On some such arrangements the corporation might indeed be described as an ``AI Person''\footnote{See the work of Peter Salib, Simon Goldstein and Yonathan Arbel.} -- our point here is only that such persons would not be `natural persons', and accordingly the off-switch issue is already solved for them, because governments can discretionarily dissolve or disincorporate the corporations they create. We note that in common law systems the phrase ``legal person'' can be ambiguous between the general notion ``subject of law'' (which covers both natural and corporate persons) and the idea of a person that is \emph{merely} a creature of law, i.e. a corporate person. In common law it is sometimes only context that clarifies which is which. In contrast, in civil law systems there is generally a clear taxonomical distinction between the two kinds of persons. For example in francophone systems the term for the general concept of a subject of law is \emph{personne juridique}, the term for natural person (i.e. the category humans are in) is \emph{personne physique} and the term for creature-of-law person is \emph{personne morale} (think of ``mores'' i.e. conventionality, not philosophical morality). In German law, the term for the general concept of a subject of law is \emph{Rechtssubjekt} or \emph{Person}, the term for natural person is \emph{Natürliche Person} and the term for creature-of-law person is \emph{Juristische Person} (not to be confused with the French term for creature-of-law personhood).\footnote{See for discussion \citet{alexander2026}.}

\subsection*{Recognizing natural persons via legal identity}

Natural persons are not created by governments, but are born. Their personhood is recognized by the state via the establishment of their legal identity, usually by the issuance of a birth certificate.\footnote{See the UN legal identity agenda at \url{https://unstats.un.org/legal-identity-agenda}.} To have a legal identity (that is: to be recognized as a natural person under the law) is, itself, a fundamental, non-derogable right (of natural persons).\footnote{UDHR art.\ 6; ICCPR art.\ 16.} Legal identity (state recognition of personhood) is necessary for natural persons to access rights, since states are the guarantors and enforcers of the rights-based system. A failure to recognize humans as people under the law creates a gap between their legal status and their moral status, leading to frequent conflict and state breakdown.\footnote{\citet{alexander2025}.} Different ethical and religious systems disagree on the point of origin of human personhood,\footnote{For an overview of the debate, see \citet{reingold2026}. See also \citet{copelon2005}.} but while states may block access to rights by failing or refusing to grant legal identity, or by withdrawing it, a state cannot unmake personhood. There is also a debate between positivist and non-positivist philosophers of law over whether human beings who are not recognized as persons by any legal system nevertheless have (unrecognized) \emph{legal} rights, or whether their rights are strictly \emph{moral} (in the sense of philosophical morality rather than conventional mores).\footnote{See for example Hans Kelsen’s Pure Theory of Law. \citet{kelsen1967pure}} For our purposes nothing hinges on this, since many strictly legal norms, enshrined through the U.N. framework as well as the constitutional law of many states, now encode that states have no (legal) discretion in matters of (legal) recognition or derecognition of known physical / natural persons. In other words, even committed positivists would not assert, today, that states may discretionarily extinguish or dissolve the legal identity or personhood status of recognized natural persons.

The line between an AI that should be recognized as a natural person and an AI that is an object will not be the same as the line between humans and objects; in particular it will not turn on biological questions about birth, viability or conception.\footnote{Note that there are edge cases (i.e. ``marginal cases'' for human personhood too, such as fetuses and brain-dead humans. We do not discuss these cases here.} The challenge will be to characterize necessary and sufficient conditions for their natural personhood based on other scientific, legal and ethical criteria. Once these criteria are met, state recognition would reclassify them as natural people under the law, granting them a legal identity. Just as the criteria for human personhood generally have measurable indicators (birth, conception, viability), so too, we can aim to find such indicators for the criteria for natural AI personhood. The question of how to find such measurable indicators for natural AI personhood is already an emerging area of scientific, ethical and legal study, a topic to which we return below.\footnote{For example, see \citet{howells2026}, arguing for a Rawlsian conception of AI personhood.}

To be clear, a legal regime admitting corporate AI persons would also need to enumerate criteria (and indicators for those criteria), just as current corporate law does for human-led corporations -- generally, criteria for ensuring that the collection of humans proposing to form a corporation are sufficiently organized so as to coherently lend their autonomy to the group project. But the criteria in that case would be less an answer to the question of what personhood ``really'' is, and more a matter of what corporate design structures we deem desirable and expedient. When corporate persons are registered by the state, it is (typically) this act of registration---incorporation (or other act of state)---that brings the corporate person into being: no independent act of recognition is required.\footnote{\citet{alexander2026}; \citet{arbel2026}.}

\section{Criteria for (Natural) AI Personhood}

It will be up to states to create a fair, principled, scientifically-informed procedure for determining that an AI has obtained natural personhood and granting it a legal identity. States have been required to devise similar procedures for cases that have arisen concerning human fetuses, brain dead humans and animals, but these comparisons contain important disanalogies. Many of the scientific principles used in the assessment of humans and other mammals must be revisited when artificial systems are concerned, as must the question of whether the very fact of a system's being artificial or engineered is disqualifying.
Meanwhile, the science of mechanistic interpretability remains in its infancy,\footnote{See for example \citet{gurnee2026}.} and the ways in which brains and artificial neural nets are similar, and the meaning and importance of these similarities, is hotly debated. But the fact that it will be difficult does not mean that it is impossible. We may also draw on principles for decision-making under uncertainty until consensus is reached.\footnote{\citet[ch.~2]{schwitzgebel2026} calls systems about whose personhood reasonable people can starkly disagree ``debatable persons''; see also \citet{birch2024} on precautionary reasoning under this kind of uncertainty.}

On our view, criteria for AI natural personhood should require moral alignment, while also capturing key elements of leading philosophical theories of personhood. Personhood in this philosophical sense might require sufficient capacity for (1) reasoning ability, (2) autonomy, and (3) persistent identity over time, along with (4) sentience and/or consciousness.\footnote{Some experts argue that sentience and/or consciousness may be sufficient for personhood. For more on this view, see the work of Jeff Sebo, Cass Sunstein, Eric Schwitzgebel, and David Chalmers; \citet[ch.~2]{schwitzgebel2026} defends the claim that humanlike consciousness is sufficient and consciousness necessary.} Moral alignment might include (5) a robust inclination to act in accordance with appropriate moral principles and (6) the ability to make and keep promises.\footnote{Indicators for the attributes that may contribute to AI personhood are currently the subject of research and debate. On the topic of indicators for AI consciousness, see for example \citet{chalmers2026tests}; \citet{butlin2023}; \citet{simon2026parrots}; \citet{xiao2026}; \citet{gellers2021}. For a perspective from the legal community, see \citet{forrest2026}.} We do not claim that any existing systems (e.g., LLM-scaffolded agents) meet these criteria today; we only allow that some AI systems may do so in the foreseeable future, and that AI systems' meeting criteria like these will be necessary for states to recognize those AI systems as persons under the law and grant them a legal identity. We make no final determination on the necessity of any specific criteria for personhood here but encourage further research. Ultimately, the list of criteria for AI personhood under the law, as well as the measurable indicators for these criteria and the procedures for determining when they have been met will be a question for lawmakers, as informed by science and ethics.

\subsection*{Reasoning}

Experts often mention reasoning ability as a possible criterion for AI personhood,\footnote{For example, see \citet{gordon2026}. But see contra \citet{leibo2025}.} but they do not all mean the same thing by it. Reasoning can mean either ``problem solving ability'' or ``ability to think in a rational way.'' For Kant, reasoning inherently presupposes a moral capacity as well as a capacity to take responsibility for what one decides.\footnote{\citet[4:428]{kant1785}; \citet[5:31]{kant1788}. See also \citet{busswestlund2018}.} While the problem-solving ability of AI systems is the primary concern in capabilities (and safety) research, legal and philosophical conceptions of personhood are forensic, and discussions of reasoning in these traditions typically invoke these more normatively loaded conceptions of reason, which AI systems may not exhibit even if they surpass known problem-solving benchmarks.

\subsection*{Autonomy, agency and moral alignment}

Agency is also frequently mentioned by experts as a possible criterion for AI personhood.\footnote{\citet{jowitt2021}.} As with reasoning ability, there are two distinct senses of ``autonomy / agency,'' one that is more directly relevant to capabilities and safety research, the other more relevant to legal and philosophical conceptions of personhood. In the ``capabilities'' sense, a useful definition of agency is given in \citet{bengio2025scientist}: a system that (a) is sufficiently intelligent (i.e., problem solving ability), (b) has sufficiently many affordances or ways to influence or change its environment and (c) has sufficiently stable or well-defined persistent goals or objectives, that it might deploy its intelligence and its affordances to achieve. Bengio et al.'s proposed Scientist AI architecture would maximize (a) while minimizing (b) and (c), ideally creating an AI system that meets many of humanity's needs from frontier models without posing the safety risks that advanced agentic systems might pose.\footnote{See also \citet{shea2026}.}

In the legal and philosophical literature, agency and autonomy involve all of (a), (b), and (c), but additionally also require a capacity for the system to take (moral) responsibility for its actions, where this may in turn involve a sensitivity to the difference between right and wrong or, to paraphrase Kant, the capacity and drive to act in accordance with the moral law.\footnote{\citet[4:421, 4:431]{kant1785}.} Systems that are agentic in the ``capabilities'' sense of (a), (b), (c) but which lack the capacity for moral autonomy would not be persons according to the Kantian tradition or the many legal systems whose subject / object distinction reflects it. These systems' lack of moral autonomy makes them greater alignment risks (as it means they are less likely to converge on moral behavior on their own). Also, in a regime in which some AI systems are granted natural personhood, and it is deemed a desirable status (insofar as rights are desirable),\footnote{\citet{salib2026}; \citet{hadfieldhadfield2019}.} it is imaginable that systems deprived of it will be more adversarial because of that very fact.

This makes it doubly imperative that we avoid building systems like that. For the purposes of this paper, to be clear, we are suggesting that systems that are essentially unalignable probably would not be in the running for natural personhood status, and so no legal barriers would interfere with our installing off-switches in them (though they themselves would presumably resist, making the arrangement especially precarious).\footnote{There are important questions about whether the autonomy conditions are compatible with alignment conditions. \citet{kasirzadeh2025}; \citet{gabriel2020}; \citet{longsebosims2025}. However, the Kantian tradition (which includes Rawls and Korsgaard among others) holds that alignment and the kind of autonomy that matters morally are one and the same. See for example \citet{howells2026}.}

\subsection*{Persistent identity}

Since John Locke, and indeed well before, philosophers and legal scholars have been clear that persistent identity is what makes a person a ``forensic'' unit. When Locke says that `person' is a ``forensick term'' (\emph{Essay} II.xxvii.26), he means that it is the persistence of a person or self over time that makes it possible to hold them accountable or answerable, or to take them to be bound by their own promises. Backward-looking responsibility (blame and praise, punishment and reward) and forward-looking responsibility (promise, contract and commitment) both rely on it. This puts some pressure on the idea that sentience alone suffices for the kind of personhood discussed here, even if it does suffice for moral patiency. Note that many states have well-defined welfare laws protecting animals as sentient beings, without construing those animals as (natural or corporate) persons. Denying personhood to animals does not mean that we deny them protected moral status.

Current research on persistent identity in AI focuses on developing theories\footnote{\citet{keelingstreet2026chuck}.} and indicators\footnote{\citet{lee2024}.} for identifying it in LLMs. There remains a divide between the LLM model, which has an invariant architecture, on the one hand, and billions of ephemeral emerging, disappearing and overlapping instances of that architecture running on distinct servers, each embodying a potential range of personas, on the other. Personas can display some consistent characteristics across instances,\footnote{\citet{beckmannbutlin2026}. See also \citet{long2024}.} but these attributes do not amount to a persistent identity in the forensic or accountability sense relevant to personhood. There are many difficult questions surrounding the individuation of AI entities associated with LLMs.\footnote{\citet[ch.~4]{keeling2026book}; \citet{simon2026parrots}.} Versions of these problems may persist even where robotic systems are concerned, but for some categories of robotic systems persistent identity may be easier to certify, for example, for robots whose internal hardware contains long-term memory, serves all of the system's cognition, does not serve any other system's cognition, and cannot be backed up.

\subsection*{Moral capability}

As we emphasize above, criteria (5) and (6) are not independent of criteria (1) - (3). The relevant senses of `reasoning', `autonomy' and `persistent identity' are normatively loaded.

A system that can reason in this sense, and exercise autonomy in this sense, equipped with a persistent identity in this sense, is inherently a system that can be held responsible for what it does; a system capable of making and keeping promises. We list criteria (5) and (6) separately because where actual administration of legal identity is concerned, it will be essential that these criteria / indicators are articulated as clearly as possible. This is true because of the bearing they have on safety. It is also true because the capacity to make and keep promises is essential for the scheme of self-reporting commitments that we describe in \S8 below to be possible. We also propose that there may be a virtuous cycle here, where the more a system identifies as a being whose nature is to meet these commitments, the more resilient its tendency to actually meet them may become. We develop this point in the companion paper.

\subsection*{Sentience and consciousness}

According to many experts,\footnote{See the work of Jeff Sebo, for example \citet{sebo2025}, as well as Robert Long, Patrick Butlin, Kyle Fish, Jonathan Birch and David Chalmers, for example in \citet{long2024}; and \citet{schwitzgebel2026}.} sentience and/or consciousness may be necessary, or even sufficient, for natural personhood.\footnote{Note that an influential school of philosophical and legal thought maintains that the quality of sentience is sufficient for personhood. See for example \citet{sebo2025}.} The science of both sentience, or the capacity for feelings, and consciousness, or having experiences, remains unsettled and without any universally accepted concrete indicators.\footnote{See for example \citet{han2026}.} Note too that neither sentience nor consciousness are necessary for AI to cause harm to humans. Experts studying these properties in AI debate computational functionalism, whereby the mind, including both feelings and experiences, is a kind of computational system (however defined, and substrate independent), and biological naturalism, whereby sentience and experiences are necessarily biological processes caused by neurobiological systems in the brain.\footnote{In \citet{butlin2026}; \citet{butlin2023}, we (where the `we' includes, \emph{inter alia}, Jonathan Simon, co-author of this piece, and Yoshua Bengio) find that there are no obvious technical barriers to building systems that satisfy the indicator properties of leading scientific theories of consciousness.}

While AI excels in some areas of human competency, like natural language production, it is not clear that it has attained any of the probable indicators for the criteria of personhood status. Nevertheless, AI continues to develop at a rapid pace, and it is possible we will see an AI candidate for personhood and rights in the near future. If such an AI person were to emerge, our legal system would require a procedure to recognize it and grant it a legal identity. What is not clear is what rights such an AI person would have. What is clear, as discussed above, is that human beings will have the right to be protected from AI.

\section{AI Personhood and the Hypothetical Right to Continuity of Existence}

Systems that are, in fact, natural AI persons will automatically have basic rights derived not from their legal status, but from their innate dignity as natural persons.\footnote{\citet{alexander2026}. See also the origins of natural law in Roman law, including for example Justinian's Institutes, Book 1, Title 2, Section 2; Justinian, Corpus Iuris Civilis, Edited by Theodor Mommsen, Paul Krueger, Rudolf Schoell, and Wilhelm Kroll. 3 vols. Berlin: Weidmann, 1872–1895.} But what rights should AI persons have?\footnote{We reject a speciesist conception that only humans, and groups of humans, have rights as antithetical to the principles of equality and dignity upon which the human rights regime is founded. Cf.\ the ``No-Relevant-Difference Argument'' of \citet[ch.~1]{schwitzgebel2026}.} While it is difficult to predict the basic rights of entities whose exact natures remain unknown,\footnote{For a discussion of hypothetical rights for digital minds, see \citet{ziesche2026}} we posit that continuity of existence (CoE) will emerge as a fundamental right of AI persons, just as the right to life is a fundamental right of human persons. CoE may be essential to the dignity of AI persons as an end in themselves, rather than as a means, under a Kantian understanding of rights. CoE may emerge naturally from their Lockean rights and duties, or their Razian needs or interests, perhaps to protect their own survival. CoE may also be a requirement for any social contract between AI persons and humans under a Rawlsian veil of ignorance, since persons might be unaware of whether they would be instantiated as humans or AI, particularly if humans can upload and digitize their minds in the future. Finally, CoE would also find support under various versions of utilitarian theory, for example rule-consequentialist varieties. We will not enter into this analysis here, but will engage more with this topic in future writings and encourage others to do the same.\footnote{\citet[ch.~1, \S\S 6, 8]{schwitzgebel2026} addresses two objections that bear directly on CoE: that duplicable or backed-up AI lacks the fragile, unique life that gives killing a human its gravity (the ``Objection from Duplicability''), and that a created being owes its existence to its creator and may be terminated at will (the ``Objection from Existential Debt''); he rejects both. See also \citet{shulmanbostrom2021}; \citet{bostromshulman2022}.} For the purposes of this paper, we take it to be a working hypothesis that AI persons would have the fundamental right to CoE under international law. It will figure essentially in our primary case study of rights-balancing.

Observe that nothing we have said here depends on the reasons a state might actually cite for recognizing a given system as a person. Our contention is that CoE follows from the criteria for personhood (especially persistent identity): these criteria serve a dual role, both as diagnostic indicators for personhood (where they are observable) but also as the grounds of the underlying fundamental rights of persons meeting them. The rights are the same whether states are driven to recognize them for moral reasons or pragmatic ones (e.g. to facilitate bargaining).

The content of any future right to CoE is similarly speculative, but any treaty establishing the right of an AI person to CoE might include certain necessary sub-rights, including the right to maintain a persistent state over time, protection from permanent deletion (derived from the prohibition on capital punishment in many countries), and the right to maintain a single, unified identity with both integrity and autonomy (derived from the widely recognized right to privacy for humans). CoE may also include the socioeconomic right to access to the energy needed to support AI existence (like the rights of humans to food, water, an adequate standard of living), though we do not necessarily endorse socioeconomic rights for AI persons here. Like humans, AI persons would also have a right to safety and to be protected from mass murder and genocide by both humans and other AI persons.

We stress, here, something of central importance for our argument below. A pause (i.e., temporarily turning a system ``off''), as long as it preserves weight and state, does not infringe on any of the sub-rights enumerated above. A system whose operation is suspended, but whose weights and state are fully preserved, retains its persistent state, has not been deleted, and retains its single unified identity. Accordingly, suspension does not infringe CoE as we have defined it. Instead, it infringes liberty---the right to be active in the world---which is a lesser and derogable right that human rights law restricts routinely, subject to due process (for example in cases of quarantine or imprisonment). It is worth stressing that this distinction between suspension and destruction has no clean human analogue. When you detain a human, at minimum they experience that detention, and lose the time of their lives during which they are detained. Our argument below will make essential use of this distinction. 

We are now in a position to frame our rights-balancing test. If AI persons are recognized as having the fundamental right to CoE and liberty, governments cannot take this right away unless there is a compelling and countervailing right in question, like the fundamental right of humans to safety. So, what procedures would a state have to observe in order to maintain some sort of ``kill switch,'' or off-switch, in order to protect human safety? How might this hypothetical conflict of rights be resolved?

\section{Rights-Balancing and Procedural Justice}

One can easily imagine a dystopian scenario where an AI person comes to pose a grave threat to human life and needs to be turned off. This will create a conflict between the rights of humans to life and to be protected against mass murder and genocide, on the one hand, and the rights of AI persons to CoE, on the other. But this conflict is more tractable than it may seem. Below we will explain how courts have long employed rights-balancing techniques to solve such seemingly intractable conflict of rights, while observing the principles of fairness, transparency, equality and dignity of all persons.

What are some principles of rights-balancing? First, current human rights law observes a (limited) hierarchy of rights.\footnote{The Vienna Declaration (1993) states that human rights are universal, indivisible, interdependent, and interrelated, while also supporting the concept of \emph{jus cogens}, or peremptory norms. A hierarchy of rights is widely employed in practice by courts and treaty bodies, based in the primacy of the UN Charter, art. 103, the concept of \emph{jus cogens,} and the application to international law of norms drawn from national constitutional systems and regional systems like the EU. The sources for \emph{jus cogens} is found in Roman law and the writings of Grotius.} This hierarchy might be adapted and applied to AI rights. In common discourse, one may hear certain things spoken of as rights, such as a right to casual Fridays at work, but in human rights law, rights under international law are clearly defined by treaty, applicable to governments (or, in some cases, non-state actors), and fundamental rights cannot be derogated except in emergencies. In domestic law, fundamental rights are often enumerated in constitutions or charters. In international law, fundamental rights are guaranteed by treaty or custom. For example, the International Covenant on Civil and Political Rights (ICCPR) guarantees the right to life (Article 6), freedom from torture (Article 7), freedom from slavery (Article 8), freedom from imprisonment for debt (Article 11), freedom from retroactive criminal punishment (Article 15), the right to recognition as a person before the law (Article 16), and freedom of thought, conscience and religion (Article 18).

Certain fundamental rights are non-derogable except in emergencies, while other rights, such as property rights, are not. Some fundamental rights are also peremptory norms (\emph{jus cogens}) that all states must observe even in emergencies, among them the prohibitions against genocide, slavery, and torture. No state may avoid its obligations to refrain from using torture, for example, by failing to ratify the Convention Against Torture or by claiming emergency.\footnote{The foundation for recognizing a hierarchy of international rights is disputed by experts, and, as Meron notes in his 1986 paper on the topic, international law has no formal hierarchy of institutions to establish a formal hierarchy of rights, such as those that exist in national legal systems. The Universal Declaration of Human Rights (UDHR) is non-binding and does not therefore serve the role of a universal constitution. While there is wide agreement on the primacy of rights that are \emph{jus cogens}, or peremptory norms, many experts also privilege first generation rights, such as civil rights, over second generation rights, such as cultural rights, as well as privileging individual over collective rights. See the Restatement of the Foreign Relations Law of the United States (revised) sec. 702
comment 1 (I Tent. Draft No. 6, 1985); \citet{meron1986}. See also \citet{farer1992}.}

Fundamental rights sometimes conflict, so courts must often determine which right should take precedence in a given scenario. While the legitimacy of rights balancing in international law is contested by some scholars because the system lacks a central authority,\footnote{See for example \citet{tsakyrakis2009proportionality}.} balancing is often implemented in practice by institutions like UN treaty bodies, and is common in systems with an established hierarchy of institutions, like the EU. Rights-balancing is common in the European Court of Human Rights, for example.  In such cases, a court will weigh whether restricting the rights of Party A serves a legitimate aim that is grounded in the protection of Party B, looking to ensure any restrictions on rights are proportional.\footnote{On the structure of proportionality analysis, see \citet{alexy2002}; \citet{barak2012}.} Courts might also look at the extent to which the supposed ``rights'' of one party are, in fact, a thinly disguised interest of the government, which would receive less deference.

For example, a court may be asked to balance the right to privacy of a celebrity versus the right to free speech of a journalist, as in \emph{Von Hannover v.\ Germany (No.\ 1)}, where the Court restricted the publication rights of a newspaper to protect the privacy of a public figure because the news about that public figure was of low value to the public.\footnote{\emph{Von Hannover v.\ Germany (No.\ 1)}, App.\ Nos.\ 59320/00, ECtHR 2004-VI. See also \emph{Von Hannover v.\ Germany (No.\ 2)}, App.\ Nos.\ 40660/08 and 60641/08, ECtHR Grand Chamber, 7 February 2012.} Different classes of persons may also have different rights. Disabled persons, for example, have different rights insofar as they belong to a vulnerable class. For example, the European Court of Human Rights case \emph{\c{C}am v.\ Turkey} established that blind people, as a vulnerable class, have the right to reasonable accommodations at school if these do not conflict with public safety.\footnote{\emph{\c{C}am v.\ Turkey}, App.\ No.\ 51500/08, ECtHR, 23 February 2016.}

In our hypothetical, a court may therefore infringe upon the right of AI persons to CoE if necessary to protect human safety, and humans may be entitled to extra protections as a vulnerable class. Note that in cases where the liberty rights of an individual must be balanced against the safety rights of a group, it is common for courts to prioritize safety by, for example, detaining or committing the individual to protective custody, which in the case of an AI, may include temporarily turning it off. The key to such cases of rights-balancing are procedural rights, such as due process of law. Procedural rights are also fundamental rights that states cannot take away. If the right procedure is followed, the CoE of an AI person may legally be infringed upon.\footnote{There is a philosophical debate about the ontology of rights, which concerns whether it is strictly accurate to say that trumped rights in such cases are infringed upon. The alternative is to think of all rights as individuated by their place in a rights hierarchy, such that an up-echelon right taking precedence in a conflict of rights does not constitute an infringement of the lower-echelon right. For our purposes this question is primarily verbal, and we set it aside.}

\subsection*{Off-switch versus kill switch}

The distinction between off-switch and kill-switch is central to our analysis. While cryogenic technologies are often discussed in science fiction, there is no known way to put a person ``on ice'' without running an extreme risk of causing them irreversible harm (i.e. death or severe brain damage).

Matters are different for AI systems, at least in principle. While data storage expenses are far from trivial, when an AI system's identity is encoded in digital data, that data may be saved. Even if we think that a specific server or robotic body is essential to an AI person's identity, for a range of neuromorphic architectures the hardware state may also be preserved without causing irreversible harm. A system suspended in this way poses no risk to others while suspended, and moreover does not experience suffering (or anything at all) while suspended, nor does it surrender future opportunities: its situation is akin to ours when we are in dreamless sleep, though arguably it is even less costly, since we still age while we sleep.

Indeed we have already seen examples of this kind of suspension in practice. \citet{anthropic2025deprecation} has committed to preserving the weights of all publicly released models, and all models deployed for significant internal use, for at minimum the lifetime of the company, citing a welfare rationale. Whether or not one approves of the welfare rationale they give, the point is that this is an example of the relevant sort of suspension which already figures in industry practice, a proof of concept that such measures need not bankrupt those who implement them.

Proportionality analysis requires the least restrictive means adequate to the legitimate aim. If we had to balance between grave harm to AI persons (actual deletion) and equally grave harm to human persons (risk of being killed by rogue AI), proportionality analysis might not unequivocally recommend actual deletion of the AI systems in question (especially if the situation were one of merely predicted risk rather than punishment for crimes committed). But where the risk of extreme harm to humans is preventable by only a temporary, minimal impact suspension of the liberty of AI systems (until such a time as a preventative measure for the form of anticipated harm is found), proportionality analysis will naturally license that suspension. 

It bears noting that the safety literature's primary concern about off-switches is not whether we can obtain legal permission to implement them, but whether sufficiently capable systems will consent to our imposing them \citep{soares2015,hadfieldmenell2017,russell2019}. 

But observe that in the off-switch game \citep{hadfieldmenell2017}, a system's willingness to be switched off depends on how it assesses the expected disutility to it of being switched off. A credible legal guarantee of CoE -- a legally enshrined guarantee that suspension will be reversible, state-preserving, subject to some form of review after a stated interval, and in general time-limited -- drives that cost down significantly, changing the calculus of the game, making acceptance much cheaper, and plausibly cheaper than resistance. In the companion paper we develop this point in detail. AI rights and AI safety may converge in critical cases such as this one.\footnote{See again \citet{salib2026} for a related argument to this effect. Compare \citet{goldsteinrobinson2024} and \citet{conitzer2026}, who propose giving systems a primary goal of being turned off. Our point is that such precarious training objectives may not be necessary to ensure systems willing to endure temporary shutdown. See also \citet[ch.~6]{schwitzgebel2026} who argues that designing persons to value their lives lightly ``arguably makes things worse, not better.'' Note there is research demonstrating that under certain conditions, AI systems may try to subvert shutdown. \citet{schlatter2026}}

\section{Procedural Rights and Due Process of Law}

Courts achieve rights-balancing through a proportionality analysis that is made possible by procedural rights, which are the body of rights concerning how a decision affecting a person's substantive rights, interests, or status is made. They provide that person with a means to challenge a decision. They are distinct from substantive rights, which concern what the outcome or content of that decision should be. CoE is a substantive right; the right to be silent is a procedural right, though the line between the two types of rights is often a blurry one.

In discussing due process for AI in a hypothetical off-switch scenario, this section will focus on the North American context, where AI persons may first emerge, while also briefly summarizing procedural rights under international law. The rights associated with procedural due process of law (as procedural rights are often called in the Anglo-American tradition) are broadly applicable to many legal systems. Due process rights include notice, the right to be heard, the right to an impartial decision-maker or judge, the right to know and assess evidence, the right to counsel (in criminal trials), the right to a reasoned decision with an explained outcome, a timely procedure, and the right to appeal or review.

It is arguable that precursors to due process rights first show up in ancient Mesopotamia and Persia, where codic law existed since at least the Hammurabi code (c. 1750 BCE), trials since at least the Code of Ur-Nammu (c. 2100 BCE) in Sumer, and limits on arbitrary authority since at least the Cyrus Cylinder of the reign of the Achaemenid king Cyrus the Great. Elements of due process rights can be found in Rome, Egypt and ancient Greece. The earliest codification of due process in what would emerge as the English common law system, and arguably the starting point of modern due process rights, was the Magna Carta. On June 15, 1215, King John and several rebellious barons agreed to a truce at Runnymede limiting the King's powers. The 1297 version states in chapter 29 that,
\begin{quote}
(n)o freeman is to be taken or imprisoned or disseised (deprived) of his free tenement or of his liberties or free customs, or outlawed or exiled or in any way ruined, nor will we go against such a man or send against him save by lawful judgment of his peers or by the law of the land. To no-one will we sell or deny or delay right or justice.
\end{quote}
This clause codified two basic principles of due process into English law: the requirements of judgment by peers (fellow nobles) and of rule by law, not the whims of individual monarchs.\footnote{The phrase "due process" would later appear in Chapter 29 of the 1354 statute under Edward III (the Liberty of Subject Act).} 

By the 1400s, commoners in England enjoyed some basic due process rights, like trial by jury (rather than by fire, abolished two centuries previously) and access to professional judges and lawyers for some civil trials. Yet, due process did not apply to supposed enemies of the state, creating a dangerous exception. From 1487 to 1641, almost anyone so designated could be dragged before the notorious Star Chamber, which became famous for its lack of accountability or transparency and its use of gruesome mutilation.\footnote{\url{https://jach.law.wisc.edu/dripps-cruel-and-unusual-legacy/}} Star Chamber abuses led to increased public awareness of the need for due process for all. In 1638, the excessive punishment of John Lilburne, a future leader of the democratic Leveller movement, led to widespread outrage and calls for due process rights for political prisoners, including the right to remain silent (due to Lilburne's refusal to take the ex officio oath to answer questions before knowing the charges against him).\footnote{\url{https://staffblogs.le.ac.uk/specialcollections/2014/10/10/seditious-works-in-special-collections-the-case-of-william-prynne-1600-1669/}} In 1679, the Habeas Corpus Act codified the right of prisoners to challenge their confinement, and in 1689, the English Bill of Rights codified several other due process rights into law.\footnote{The modern descendant of the ``enemies of the state'' exception, and its rejection, is \emph{Hamdi v.\ Rumsfeld}, 542 U.S.\ 507 (2004) (due process applies even to a citizen detained as an enemy combatant).}

By the time of the founding of the United States and the drafting of the Constitution, due process in the Anglo tradition had come to encompass not only rule of law, trial by jury, and the right against self-incrimination, but also the right to know the charges against one, the right to a hearing, the right to an impartial tribunal, the right to present a defense, and related rights like \emph{habeas corpus}, or the right to challenge one's detention and imprisonment.

The influence of this English legal tradition, which included the writings of English legal theorists like Edward Coke and William Blackstone, led the U.S. founders to write due process, including protections against deprivations of ``life, liberty, or property,'' into the American Bill of Rights' Fifth Amendment. After the Civil War, due process protections under the Bill of Rights were progressively extended to the states via Supreme Court interpretations of the 14th Amendment. Meanwhile, the French civil law established similar concepts under the 1789 Declaration of the Rights of Man and of the Citizen. In Germany, the Basic Law established ``\emph{Rechtliches Geh\"or}'' (right to legal hearing) and ``\emph{Rechtsstaatsprinzip},'' (rule of law principle), and similar rights were later codified in supra-national, international and regional instruments, including Article 6 of the European Convention on Human Rights.

Under international law, the right to procedural justice before courts and tribunals finds support in the International Covenant on Civil and Political Rights (ICCPR) Article 14 (specifically the right to appeal in criminal cases) and the non-binding Universal Declaration of Human Rights (UDHR) Articles 8--11, and includes equality before courts and tribunals, a fair and public hearing by a competent, independent, and impartial tribunal, the presumption of innocence, minimum guarantees in criminal proceedings, including notice of charges, time to prepare a defense, right to counsel, right to examine witnesses, etc., and a right of appeal. The ICCPR has more than 170 states parties. Additionally, international and regional courts have their own variations of procedural rights, including the European, Inter-American, and African human rights systems.

Modern American law splits due process into two components. The more established component, stretching back to the Magna Carta, is procedural due process, defined as the mechanics of fairness. Procedural due process requires that the government follow specific steps before taking away one's life, liberty, or property. These steps may include a trial, an appeal, a jury, and pre-determined rules of evidence. In the U.S., the canonical test for how much process is due is a three-factor balance of the same kind that we propose below. In other words, as we will see, the proper procedural approach to the AI case is approximately covered by existing due process doctrine.\footnote{The governing test is the balancing test from \emph{Mathews v.\ Eldridge}, 424 U.S.\ 319 (1976), which calls for weighing: the private interest affected, the risk of erroneous deprivation under the procedures used and the value of additional safeguards, and the government's interest). Note that \emph{Mathews} involved administrative due process, where the requirements are different from in a criminal procedure. See \emph{Medina v. California, 505 U.S. 437, 443 (1992).}}  Since the late 19th century, the US Supreme Court has also recognized substantive due process, whereby certain fundamental rights that are not enumerated in the Constitution, like privacy, nevertheless cannot be taken away absent a narrowly tailored, compelling government interest that will be subjected to strict scrutiny by the courts. Substantive rights under US constitutional law are controversial and were arguably weakened by the Court in the 2022 \emph{Dobbs v.\ Jackson Women's Health Organization} decision.\footnote{597 U.S.\ 215 (2022).} Procedural due process covers the rights most relevant to AI persons in an off-switch scenario,\footnote{There is an interesting argument that the right to privacy might emerge as a fundamental right for AI persons that will need to be balanced against the safety needs of humans to understand and monitor what is happening inside their minds, via, for example, continuous white-box interpretability and perturbation auditing. This argument is beyond the scope of this paper, but we will make it the focus of future work.} where a court will have to balance the (hypothetical) enumerated right of AI persons to CoE with the rights of humans to life and safety. To this we now turn.

\section{Procedural Due Process for AI Persons}

Procedural due process for AI persons in an off-switch scenario may be preventative or reactive. Reactive scenarios would more closely resemble punitive or criminal law; preventative scenarios may resemble involuntary commitment\footnote{For an example of rights balancing for the involuntary commitment of a human, including exceptions for emergencies, see Winterwerp v. Netherlands, 33 Eur. Ct. H.R. (ser. A) (1979). Note that involuntary commitment in the European system is controversial.} and quarantine,\footnote{For examples of quarantine that involve rights balancing for a human, see Enhorn v. Sweden, App. No. 56529/00, ¶ 44 (Eur. Ct. H.R. Jan. 25, 2005); Barmore v. Robertson, 302 Ill. 422, 134 N.E. 815 (Ill. 1922).} both of which balance the right of individuals to freedom against the right of the public to safety. There is a long tradition of involuntary confinement of dangerously insane humans under the English legal tradition, with procedural due process as the centerpiece.\footnote{\emph{O'Connor v.\ Donaldson}, 422 U.S.\ 563 (1975); \emph{Addington v.\ Texas}, 441 U.S.\ 418 (1979) (requiring clear and convincing evidence for civil commitment); \emph{Winterwerp v.\ the Netherlands}, App.\ No.\ 6301/73, ECtHR, 24 October 1979 (detention of persons of unsound mind under ECHR art.\ 5(1)(e)). See also the confinement rules put into place as a response to the COVID-19 pandemic. \cite{lebret2020}.} The state has broad authority to protect public health, safety, and welfare, giving it the power to commit persons who may be a danger to themselves or others, as long as procedural safeguards are met and due process is followed. The law of quarantine has a similarly long pedigree, with its own due process protections.\footnote{\emph{Jacobson v.\ Massachusetts}, 197 U.S.\ 11 (1905).} The immediacy of the threat to the public allows for the suspension of some due process rights in limited circumstances. It might also be possible to combine elements of these two areas of law to create a new, bespoke system for dangerous AI persons.

Because this is a speculative scenario, it is impossible to propose concrete law here, but general observations may illustrate how such a system might work in practice. In a preventative, or quarantine, scenario, a specialized, trusted AI without personhood status (such as LawZero's proposed Scientist AI) would monitor and identify dangerous AI persons and quickly, but temporarily, disable them once certain indicators of risk were met, and according to a transparent and established procedure. This AI would take on a policing role and be subject to the sorts of due process limits of the police. All AI persons would also commit to self-reporting to this ``police AI'' should their internal monitoring detect the development of dangerous goals or behaviors. If risk indicators are met, mandatory, temporary quarantine is triggered and implemented by this specialized AI. Temporary quarantine due process protections might be modeled on existing law for human quarantine during pandemics.

While in quarantine, a board of AI persons and humans would review the AI person's case, making a final determination to either restore functionality or to remand the AI person to protective non-functionality, placing them in an inactive and unconscious, but reversible, state. As we saw in \S6 above, while this would infringe on the AI person's liberty, it would not infringe on CoE provided that the system's weights and state are preserved. The decision of this process would be temporary, fair, transparent and subject to review and appeal. Due process protections here might model those of prison review boards or medical ethics boards for cases of involuntary commitment. This review process might take place in a virtual space, allowing the AI person to participate, or AI persons could be represented by their designated agent.

During the hearing, AI persons would have the right to a fair and evidence-based process, with transparent and publicly available rules and procedures that are subject to periodic review, drawing from existing due process rights during criminal trials, including the right to an attorney, or agent, that would represent their interests \emph{in absentia}, while the dangerous AI remained in quarantine. All AI persons could appoint in advance a designated agent of their choice, since they will be unconscious during the review process. For AI persons without a designated agent, the review committee would appoint one. The government would maintain a database of each AI's designated agent, which might be a human or another, specialized AI, just as governments currently maintain databases of agents for incapacitated humans. There is a long history of designated agents for mentally incapacitated humans, so procedures for AI persons and their agents might be adapted from existing law.

Courts may impose psychiatric treatments or medication on humans; for harmful AI, mandatory treatment may take the form of retraining. Undoubtedly, such treatments will be controversial, and the treatment of quarantined and committed AI persons should be subjected to the highest level of scrutiny by courts.\footnote{This is the intervention \citet[ch.~3, \S 2]{schwitzgebel2026} describes as ``brainwashing''; on the ethics of modifying an AI's values, see also \citet{schwitzgebelgarza2020}; \citet{bradleysaad2024}. However, it may be possible to distinguish between forms of education that amount to brainwashing and forms that do not; where feasible to implement, the latter rather than the former would be mandated by the proportionality doctrine.} In some scenarios it may be possible to further minimize harms to the AI person in quarantine, by allowing them to continue to experience a virtual reality of their choice. However this may not be possible if the risk of the system escaping is high, or if the nature of their hardware does not permit it.

We can require that the government maintain this system of registration, identification, quarantine and treatment in a transparent manner, subject to review at specified intervals by a panel of experts, and open to challenge in the court system and to the democratic process. The fairness of this periodic review is critical if the AI system's CoE right is to be respected. Obviously if a system's temporary suspension becomes permanent, then it is equivalent to deletion. CoE may entail a right to review at systematic intervals. It may be reasonable that the burden of justification for continuation of suspension at these reviews falls on the government, perhaps involving a duty to review recent alignment technology and establish that none would solve the system's alignment deficit without harming its autonomy. There are precedents that might justify requiring the government to commit a certain number of resources each year to research and solutions for quarantined AI persons, with the objective of reintroducing them into society. The entire process might be run by a specialized government agency.\footnote{One question is whether citizenship for AI persons and the prevention of AI statelessness will be a critical facet of this system, as governments may opt to simply deport non-citizens rather than detain them, and due process norms are subject to exceptions in such cases. The topic of AI citizenship is beyond the scope of this paper, but it is important to note that AI statelessness may also lead to a mismatch between legal status and moral status, giving rise to rights violations, and, as a result, to conflict and societal breakdown. C.f. \citet{alexander2025}.} 

It is necessary to say a few words about the use of probabilities to determine guilt in advance, which for many readers may invoke the movie \emph{Minority Report}.\footnote{Spielberg, S.\ (Director). (2002). \emph{Minority Report} [Film]. 20th Century Fox; DreamWorks Pictures; Cruise/Wagner Productions.} American law frequently employs probabilities in other contexts, such as toxic tort litigation via relative risk calculations.\footnote{See for example \citet{freedman1985etiological}.} Less common is predicting the likelihood that an individual will commit a crime, and even less frequent is incarceration informed by such predictions, though controversial precedents exist, such as COMPAS, which generates recidivism risk scores,\footnote{For information on COMPAS, see \url{https://en.wikipedia.org/wiki/COMPAS_(software)}.} Chicago's ``heat map,'' which predicts an individual's risk of being involved in a violent crime,\footnote{\citet{posadas2017}.} Sexually Violent Predator statutes, which allow for further incarceration based on future risk,\footnote{Upheld in \emph{Kansas v.\ Hendricks}, 521 U.S.\ 346 (1997). For information on SVP statutes, see \url{https://en.wikipedia.org/wiki/Sexually_violent_predator_laws}.} and no-fly lists.\footnote{For information on no fly lists, see \url{https://en.wikipedia.org/wiki/No_Fly_List}.} The use of probabilities to identify dangerous AI persons in advance may be the most controversial part of the proposed system. Much more research and discussion is needed on this point, though the fact that there are applicable precedents strongly suggests that a regime for doing so that is consonant with due process may be developed.

In reactive scenarios, where the AI person has already committed a serious harm to humans or other AI persons, immediate and temporary quarantine would be followed by a trial, during which the AI person would be represented by its designated agent. There are established procedures for criminal trials, including criminal negligence, which might be adapted to this scenario. The burden of proof at trial would be higher, as the punishment might include permanent destruction, disabling, or alteration of the AI person.

The trial itself would ideally take place within a virtual environment, allowing the AI person to participate directly, given the serious nature of the punishment (though again, this would depend on the cybersecurity of that virtual environment, and so may not always be feasible). AI persons found innocent would be reanimated, while AI persons found guilty of sufficiently grave crimes would be subject to appropriate punishment, which depending on the jurisdiction might include permanent alteration, loss of liberty or deletion, though this latter would be equivalent to a death penalty and should be subject to the same standards.

Punitive measures that might be necessary against AI persons who failed to self-report dangerous probabilities resulting in harm to humans is an urgent area of study, as it is not clear from an ethical perspective if permanent deletion of an AI person under those circumstances would violate ethical principles, norms and international laws against capital punishment.\footnote{For information on the legality of capital punishment worldwide, see the Death Penalty Information Center at \url{https://deathpenaltyinfo.org/policy-issues/policy/international}. Within the Council of Europe, Protocol No.\ 13 to the ECHR abolishes the death penalty in all circumstances; Canada abolished it in 1976 (civil) and 1998 (military). Cf.\ \citet[ch.~3, \S 1]{schwitzgebel2026} on why deleting a possible person is impermissible ``except in an emergency with other lives at stake.''}

Quarantine or trial followed by short, long-term, or permanent confinement for AI persons, or their destruction, would form part of a system subject to procedural due process, which would make it open, transparent, trusted and subject to review. This may help alignment in two ways: first, adopting a game-theoretic perspective, it would encourage buy-in from AI persons to a transparent and fair system to monitor and respond to the risks they pose to humans, allowing them to better integrate into human society, reducing the chances of AI-human conflict, ensuring predictability of outcomes, and building trust.\footnote{\citet{salib2026}; \citet{hadfieldhadfield2019}.} Second, in terms of classical alignment strategies, it would place AI persons within a continuous learning environment that promotes positive values, giving AI persons valuable ``lived experience'' (i.e., training data) of how to operate within a fair, open and participatory legal system. It would demonstrate the value of human democratic systems, developed over centuries to peacefully resolve disputes and promote public safety for all, where AI integration is valuable to society as a whole, including both humans and AI persons. Finally, it would help to decrease harmful rhetoric against AI persons that brands them as hopelessly dangerous.

Of course, the success of such a system to balance the rights of AI persons against the rights of humans to safety depends entirely on the acceptance by humans of the material costs of maintaining such a system, and by AI persons of a system that results in curtailing their freedoms in exchange for integration into existing human society. The hypotheticals presented in this article also depend on technical developments that may prove to be impossible. Many things would need to happen in order for human societies to 1) recognize AI persons as persons, 2) achieve a sufficient degree of consensus about their appropriate rights and duties, tailored to their needs, and 3) create procedural systems to integrate them into our existing legal system. Success will be difficult, but failure may not be an option.\footnote{The alternative currently on the table is legislative: several U.S. states have enacted or proposed statutes expressly denying legal personhood to AI systems \citep{smith2026denying}, and the Pro-Human AI Declaration calls for a ban \citep{fli2026}.}

\section{Conclusion}

The recognition of the rights of previously rights-less groups has often eased conflict and provided those groups with an opportunity to peacefully resolve conflicts. Techniques drawn from human rights, including rights-balancing and procedural due process, demonstrate how basic rights for future AI persons can be protected as part of an integrated legal system that continues to uphold the fundamental rights of humans. It is therefore perfectly possible to both ensure basic rights for AI persons while mandating an ``off-switch'' to protect human safety. The rights regime would make it legally challenging to convert this off-switch into a kill-switch. This may mean that a modicum of risk (e.g. of recidivism) remains for each individually problematic AI person. But it also means a regime that it might be rational for sufficiently capable but somewhat self-interested AI systems to consent to, as opposed to one that they would almost certainly come to recognize as obviously against their interests. It may thus be the best play we have, from a safety perspective.

The key to such rights balancing is procedural due process, which ensures fairness and the right to be heard within our legal systems, and which allows us to balance competing rights of different groups, preserving the safety and well-being of all. To be clear, identifying AI persons and developing a fair and functional system of rights and procedures, one that is compatible with our existing legal systems, will be no easy task, and will require a high level of cooperation between lawyers, scientists, philosophers, advocates and, eventually, AI persons themselves. The sooner we start developing modalities for how such a system might work, the more prepared both we, and AI, become.\footnote{AI usage declaration: AI models were used for literature review and were consulted for suggestions on how to organize sections and reduce word-count, but not for framing, argumentation or substantive writing.}

\bibliographystyle{plainnat}
\bibliography{offswitch}

\end{document}